\documentclass[a4paper,11pt]{article}
\usepackage{pos} 

\title{The 't Hooft-Veneziano limit of the Polyakov loop models}

\author*[a]{O.~Borisenko}
\author[b]{V.~Chelnokov}
\author[a]{S.~Voloshyn}

\affiliation[a]{Bogolyubov Institute for Theoretical
Physics, National Academy of Sciences of Ukraine,\\
  14-b Metrolohichna str. Kyiv, 03143, Ukraine}

\affiliation[b]{Institut f\"ur Theoretische Physik, Goethe-Universit\"at
Frankfurt,\\
60438 Frankfurt am Main, Germany}

\emailAdd{oleg@bitp.kiev.ua}
\emailAdd{chelnokov@itp.uni-frankfurt.de}
\emailAdd{billy.sunburn@gmail.com}

\abstract{The broad class of U(N) and SU(N) Polyakov loop models on the lattice 
are solved exactly in the combined large N, Nf limit, where N is a number 
of colors and Nf is a number of quark flavors, and in any dimension. 
In this 't Hooft-Veneziano limit the ratio N/Nf is kept fixed. We calculate 
both the free energy and various correlation functions. The critical behavior 
of the models is described in details at finite temperatures and non-zero baryon 
chemical potential. Furthermore, we prove that the calculation of the N-point 
(baryon) correlation function reduces to the geometric median problem in 
the confinement phase. In the deconfinement phase we establish an existence 
of the complex masses and an oscillating decay of correlations in a certain 
region of parameters.}

\FullConference{%
 The 38th International Symposium on Lattice Field Theory, LATTICE2021
  26th-30th July, 2021
  Zoom/Gather@Massachusetts Institute of Technology
}

\begin{document}
\maketitle

\section{Polyakov loop models } 

The (traced) Polyakov loop (PL) on the lattice $L^d\times N_t$ is defined as  
\begin{equation}
	W(x) \ = \ {\rm Tr} \ \prod_{t=1}^{N_t} \ U_0(x,t) \ , 
	\label{pl_def}
\end{equation}
where the product runs in the compactified time direction and $U_0(x,t)\in G=U(N), SU(N)$ is the zeroth component of the lattice gauge field. 
Expectation value $\langle W(x) \rangle$ of the PL serves as an exact (approximate) order parameter of a deconfinement phase transition in pure (with dynamical quarks) gauge theory. The two-point correlation function can be related to a potential 
between static quark--anti-quark pair, while $N$-point correlation gives $N$-quark (baryon) potential. 

One possible way to reveal a phase structure of $(d+1)$-dimensional QCD is to first construct a $d$-dimensional effective model for the PL and, second, to study phase transitions in this effective model. This approach has been described in some details in \cite{philipsen_rev_19}. There are many PL models that can be found in the literature. They range from a relatively simple local model of the form 
\begin{eqnarray} 
S  =  \beta \ \sum_{x,n} \ {\rm Re} W(x) W^*(x+e_{n}) + 
\sum_x \sum_{f=1}^{N_f} h(m_f) \left ( e^{i\mu} W(x) + e^{-i\mu} W^*(x) \right ) 
\label{action_local}
\end{eqnarray} 
to a complicated non-local model of Ref.\cite{greensite16} 
\begin{eqnarray} 
S  =  \sum_{x,y} \ {\rm Re} \ W(x) \ K(x-y)  \ W^*(y) + 
\sum_x \sum_{f=1}^{N_f} h(m_f) \left ( e^{i\mu} W(x) + e^{-i\mu} W^*(x) \right ) \ . 
\label{action_nonlocal}
\end{eqnarray} 
Here, $\beta=\beta(g^2)$, $h(m_f)$ is a function of the quark mass $m_f$ and $\mu$ the quark chemical potential. An exact dependence of the effective coupling $\beta$ on 
the gauge coupling $g^2$ and dependence of the external field $h$ on the quark mass are not important here.

Our goal is to study these and similar PL models in the combined large $N, N_f$ limit called the 't Hooft-Veneziano limit at finite temperature and non-zero baryon chemical potential \cite{Hooft_74,Veneziano_76}: $g\to 0, N\to\infty, N_f\to\infty$ such that the product $g^2 N$ and the ratio $N_f/N=\kappa$ are kept fixed. For the case of $N_f$ degenerate flavors considered here one has: 
\begin{equation} 
 \sum_{f=1}^{N_f} h(m_f) = N_f h(m) \to N \kappa h(m) \equiv N \alpha \ . 
\label{alpha_def}
\end{equation}
All details of our calculations can be found in Refs.\cite{largeN_sun,largeN_sun_I}. 

In Sec.2 we outline our approach to the construction of the 't Hooft-Veneziano limit. Sec.3 is devoted to a description of the phase diagram for $U(N)$ and $SU(N)$ PL models. Various correlation functions of PLs and the screening masses are calculated in Sec.4. In Sec.5 we list our main conclusions.

\section{Construction of the large $N, N_f$ solution } 

The method used to obtain an exact solution in this limit does not rely on the large $N$ factorization and applies equally well to all PL models if the effective action depends only on the fundamental and/or adjoint PLs. 
First, insert the following unity in every lattice site 
\begin{equation}
\int \rho d\rho \int_0^{2\pi} \frac{d\omega}{2\pi} \ \delta \left (N \rho \cos\omega -  \mbox{Re} \ 
W(x) \right ) \ \delta \left (N \rho\sin\omega -  \mbox{Im} \ W(x) \right ) \ . 
\label{change_var}
\end{equation}
This can be considered as a change of variables. Using an integral representation for the delta-functions and exchanging the order of integration lead to the calculation of the group integral in the large $N$ limit 
\begin{equation}
{\cal{I}}(a,b) \ = \ \int dW \exp\left [ a W  +  b W^*  \right ]  
\label{group_int}
\end{equation}
with $a,b$ - complex parameters. 
The crucial observation we did in \cite{largeN_sun} is that at non-zero chemical potential $SU(N)$ integrals  differ from $U(N)$ ones, so that $U(N)$ integrals cannot be used to study the finite-density behavior. Actually, this observation was our main motivation to re-examine the large $N$ limit of PL models at non-vanishing chemical potential.  
We expand the integral (\ref{group_int}) into a sum over partitions as 
\begin{equation} 
	{\cal{I}}(a,b) \ = \ \sum_{r=0}^{\infty} \sum_{q=-\infty}^{\infty} 
	\ \left ( a b \right )^r \ 
	\sum_{\lambda \vdash r} \ \frac{d(\lambda)d(\lambda + |q|^N)}{r!(r+N|q|)!} \ c^{|q| N} \ , 
\end{equation}
where $c=a$ if $q>0$ and $c=b$ if $q<0$. 
$\lambda = (\lambda_1\geq \lambda_2\geq \cdots \geq \lambda_N\geq 0)$ is a partition of $r$ and $d(\lambda)$ - dimension of the symmetric group $S_r$. 
This sum over partitions can be evaluated in the large $N$ limit and leads to 
the following representation for the partition function 
\begin{equation}
Z = \prod_x \ \int_0^1\rho(x)d\rho(x) \int_0^{2\pi} \frac{d\omega(x)}{2\pi}
\int_{-\infty}^{\infty} \ du(x) \ e^{N^2 S_{eff}} \ .
\end{equation}
The effective action is given by ($W(x)=N\rho(x) e^{i\omega(x)}$, 
$\alpha = \kappa h(m)$)
\begin{eqnarray}
S_{eff} &=& S_g(\{ \rho(x) e^{i\omega(x)}, \rho(x) e^{-i\omega(x)} \} ) +
\alpha \ \sum_x \rho(x) \cos\omega(x)  \nonumber  \\   
&+& \mu \sum_x u(x) + \sum_x \  V(\rho(x),\omega(x),u(x))  \ , 
\end{eqnarray}
\begin{equation*} 
V \ \sim \ \lim_{N\to\infty} \ N^{-2} \ \ln {\cal{I}} \ . 
\end{equation*}
The first term in the effective action describes contribution from the pure gauge action, next two terms - contribution from the static quark determinant (second term in (\ref{action_local}),(\ref{action_nonlocal})). Finally, the last term arises after group integration - it can be considered as a Jacobian of transformation 
to new variables $\rho, \omega$. Important is that this term scales like $N^2$ at large $N$. Its explicit expression is given in \cite{largeN_sun_I}.

\section{Phase diagram}  

Let us describe now the phase structure of the PL model. 
The partition function and all observables are calculated by a saddle-point method.  
We look for translation invariant solutions of saddle-point equations. 

In the pure gauge case, $\alpha=0$, one finds a 1st order phase transition. 
The expectation value of the PL jumps from zero to $1/2$ at the critical point $b=1$. 
Here, $b=d \beta$ for the local action (\ref{action_local}) and $b=K(0)$ for the non-local action (\ref{action_nonlocal}), where $K(0)$ is the zeroth mode of the kernel $K(x-y)$.

When quarks are added, $\alpha = \kappa h(m) \ne0$, this turns into a 3rd order phase transition of the Gross-Witten-Wadia type \cite{gross_witten,wadia} along the critical line 
\begin{equation}
	b + \alpha \ = \ 1 \ . 
	\label{crit_line_un}
\end{equation}
This fully agrees with mean-field solutions of the large $N$ limit of $U(N)$ models 
obtained earlier in Refs.\cite{damgaard_patkos,christensen12}. It is important to stress that the large $N$ limit of $U(N)$ models does not depend on the chemical potential: dependence on $\mu$ drops out both from the partition function and from all invariant observables \cite{christensen12}.

\begin{figure}[htb]
	\centering{\includegraphics[scale=0.3]{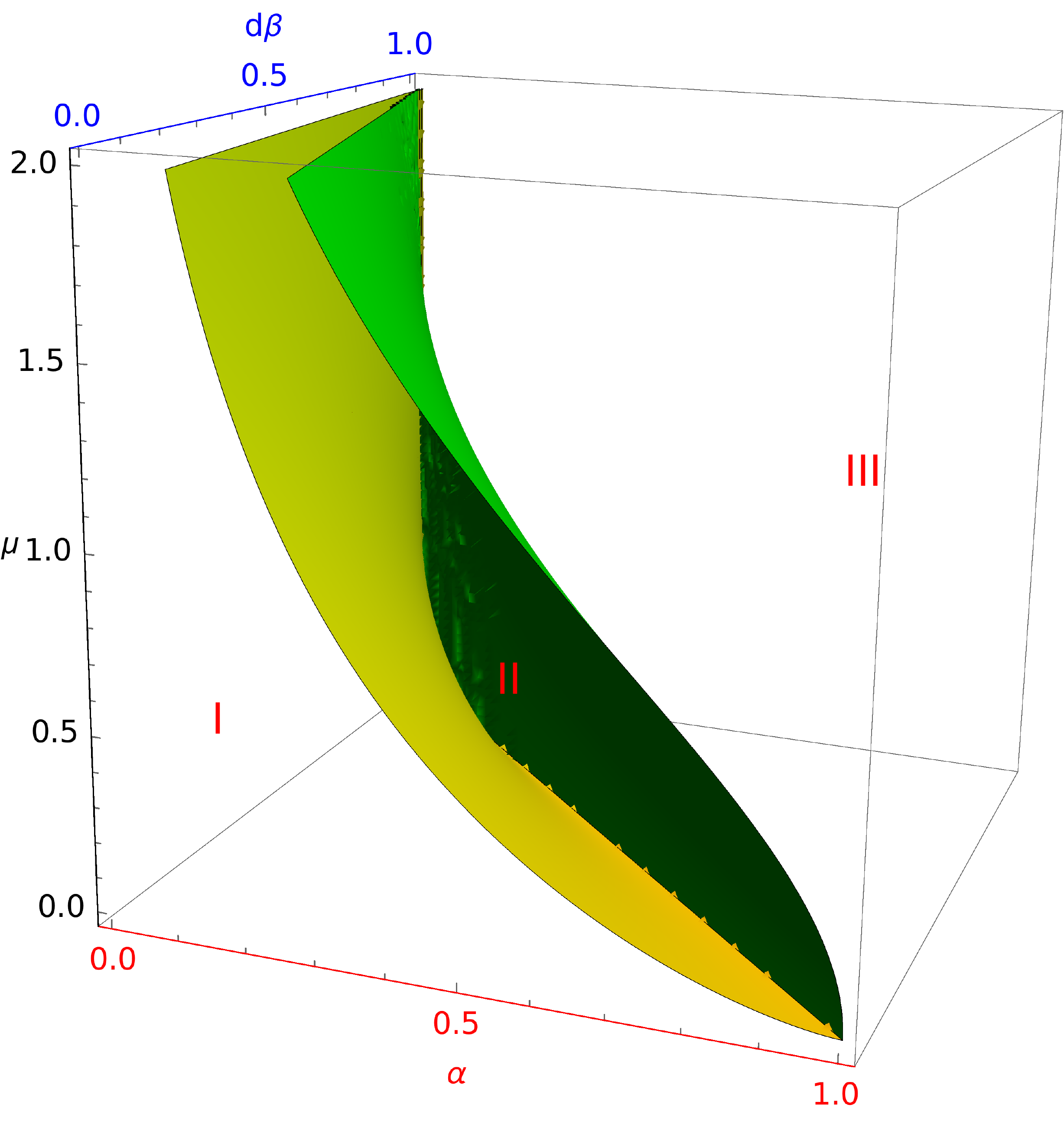}}
	\caption{Phase diagram of the PL model in the large $N$ limit. 
		Region I: no dependence on $\mu$, $SU(N)$ free energy coincides with $U(N)$ one.
		Mass spectrum is real. Region II: non-trivial dependence of the free energy 
		on $\mu$. Non-zero particle density and complex masses. 
		Region III: Non-zero particle density. Masses are real.} 
	\label{fig1:phase_diagram}
\end{figure}

The phase structure of the $SU(N)$ PL models at non-zero $\mu$ has the following form. 
We find three different regions of the phase diagram shown in Fig\ref{fig1:phase_diagram}. 

The free energy in the region I does not depend on the chemical potential and coincides with the $U(N)$ free energy. A non-trivial dependence on $\mu$ and a non-zero particle density appear in the region II. 
The critical (yellow) surface is described by the following equation
\begin{equation}
	\mu \ = \  
	\ln \left ( 1 + \sqrt{1 - z^2} \right ) - \ln z - \sqrt{1 - z^2} \ , \ 
	z = \frac{\alpha}{1-b} \ . 
	\label{crit_line_sun}
\end{equation} 
One finds a third order phase transition across this critical surface: while first two derivatives of the free energy are continuous on the surface the third derivative exhibits an infinite jump. The most interesting feature of this region is an appearance of complex masses as will be explained in the next section. 
The transition from the region II to the region III along the green surface is not 
a genuine phase transition but rather a smooth crossover. In the region III masses are real again. The green surface is approximately given by the equation 
\begin{equation} 
z\approx \cosh^{-1}\mu \ . 
\label{green_surface}
\end{equation}

\section{Correlation functions and screening masses} 

In this section we  discuss the behavior of the correlation functions of PLs 
\begin{eqnarray}
\Gamma(\eta,\bar{\eta}) = 
\langle \ \prod_x \ \frac{1}{N^{\eta(x)+\bar{\eta}(x)}} \ W(x)^{\eta(x)} W^*(x)^{\bar{\eta}(x)} \ \rangle \ . 
\label{corr_def}
\end{eqnarray} 
The correlation of an arbitrary form is evaluated by integrating over Gaussian fluctuations around the saddle-point solution. 
At large but finite $N$ the properties of $\Gamma$ depend on the dimension and 
the presence/absence of the external field $\alpha$. 
Below we consider two examples of correlations in dimension $d=3$. 

The first one deals with $N$-point function in the pure gauge theory in the confinement phase. Such correlation is related to the potential between $N$ static quarks (baryon potential) and is given by 
\begin{equation}
	\Gamma_N(\sigma)\ \sim \ \sum_x \ \prod_{i=1}^N \ G_{x,x(i)}(\sigma) \ , \ 
	\sigma \ = \ \sqrt{\frac{2}{\beta}(1-d\beta)} \ ,  
	\label{Npoint_corr}
\end{equation}
where $x(i)$ are positions of $N$ static quarks, the sum over $x$ runs 
over all lattice sites and $G$ is the Green function for massive scalar field 
\begin{equation} 
	G_{x,x^{\prime}} = \frac{\mbox{const}}{R^{\frac{d}{2}-1}} \ K_{\frac{d}{2}-1} (\sigma R) \ , \ 
	R^2=\sum_{n=1}^d (x_n-x_n^{\prime})^2 \ , 
	\label{green_func}
\end{equation}
where $K_n(x)$ is the modified Bessel function of the 2nd kind. 
Calculation of $\Gamma_N(\sigma)$ in the continuum reduces to the well-known geometric median problem: one has to find a point $y$ which minimizes the expression  $\sum_{i=1}^N\sqrt{\sum_{n=1}^d(y_n-x_n(i))^2}$. If such point $y$ is found the $N$-quark potential takes the form of the geometric median law 
\begin{equation}
	V_N(\sigma) \ \sim \ \sigma \ \sum_{i=1}^N |y-x(i)|  \ . 
	\label{GM_law}
\end{equation}
If $N=3$ this gives a famous $Y$ law for the three-quark potential
\begin{equation}
	V_3(\sigma) \ \sim \ \sigma Y \ . 
	\label{Y_law}
\end{equation} 
The quantity $\sigma$ given in Eq.(\ref{Npoint_corr}) is a string tension of the $N$-quark system. It equals the quark--anti-quark string tension. 
We think it is interesting to see how an analog of the $Y$ law appears 
for large number of colors.

Our second example deals with complex masses that appear in the region II of the phase diagram discussed in the previous section. The connected part of the two-point correlation of PLs in regions II and III of the phase diagram appears to be
\begin{equation}
	\langle \ W(0) W^*(R) \ \rangle_c = M M^* (G_R(m_1) + G_R(m_2)) \ . 
	\label{2point_corr}
\end{equation}
Here, $M$ is magnetization, $G_R(m_i)$ are diagonal correlators in the correlation matrix at $\mu\ne 0$. If $\mu=0$, $m_{1,2}$ correspond to chromo-electric and chromo-magnetic masses. If $\mu\ne 0$, in the region II masses are complex: $m_1=m_2^*=m_r+i m_i$. This leads to an exponential oscillating decay of the correlations 
\begin{equation}
	\langle \ W(0) W^*(R) \ \rangle_c \ \sim \ e^{-m_r R} \ \cos m_i R  \ . 
\end{equation}
In the region III the masses are real again with $m_2>m_1$. No phase transition separating regions II and III has been found. 
Phase with an oscillating decay was shown to exist in $(1+1)d$ LGT with heavy quarks 
in Ref.\cite{ogilvie2016} and in $Z(3)$ spin model in a complex external field in Ref.\cite{forcrand_z3}.

\section{Summary}

In this paper we have presented a qualitative description of our main results related to the 't Hooft-Veneziano limit of some PL models. Details of all calculations and full expressions for many quantities appearing in the text can be found in \cite{largeN_sun_I,largeN_sun_corr}. 
Three of the most essential results are the following: 

\begin{itemize}
	
\item 
Derivation of the large $N$ representation for the partition and correlation functions.  
\item 
Establishing the phase diagram of the $SU(N)$ model in $(\beta,\alpha,\mu)$ 
coordinates and calculation of the equation for the critical surface.
\item 
Computation of screening masses above the critical surface and an establishing 
of the exponential decay of the PL correlations modulated by a cosine function. 

\end{itemize} 

We have not found an onset transition at finite chemical potential. 
In order to see such transition one has to work presumably with an exact static quark determinant like in Ref.\cite{philipsen_quarkyon_19}. 
Such investigation is now in progress. 
Also, we think the problem of the complex mass generation at finite quark density deserves further thorough investigation

\acknowledgments

O. Borisenko acknowledges support from the National 
Academy of Sciences of Ukraine in frames of priority project 
"Fundamental properties of matter in the relativistic collisions 
of nuclei and in the early Universe" (No. 0120U100935). 

Author V. Chelnokov acknowledges support by the Deutsche
Forschungsgemeinschaft (DFG, German Research Foundation) through the CRC-
TR 211 ’Strong-interaction matter under extreme conditions’ – project number
315477589 – TRR 211.

\end{document}